\documentclass[iop]{emulateapj}
\usepackage{graphicx}
\usepackage{bm}

\def\gsim{\mathrel{\rlap{\lower 4pt \hbox{\hskip 1pt $\sim$}}\raise 1pt
\hbox {$>$}}}
\def\lsim{\mathrel{\rlap{\lower 4pt \hbox{\hskip 1pt $\sim$}}\raise 1pt
\hbox {$<$}}}

\begin{document}
\title{Merge or survive: Number of Population III stars per minihalo}
\author{Hajime Susa}
\affil{Department of Physics, Konan University, Okamoto, Kobe, Japan}
\email{susa@konan-u.ac.jp}
\begin{abstract}
The formation process of Population III (PopIII) stars in the mass accretion phase is investigated by numerical experiments.
The barotropic relation of primordial gas and artificial stiffening of the equation of state in very dense regions 
($> 10^{15}~{\rm cm}^{-3}$) enables us to follow the fragmentation of PopIII circumstellar disks and the merging processes of the fragments. The disk becomes gravitationally unstable to fragmentation 
, followed by a rapid merger process typically within 100 yrs, which roughly corresponds to one orbital time of the circumstellar disk. 
We also find that the fragmentation of the gas disk around a multiple system, a circumbinary disk, is rare; however, it is frequent in the disk around an individual protostar.
We also perform a simulation with standard sink particles, where the number and total mass of sink particles are in rough agreement with those of the stiff equation of state runs.
Based on the results of these numerical results, we model the evolution of the number of fragments with a simple phenomenological equation.
We find that the average number of fragments is roughly proportional to $t^{0.3}$, where $t$ is the elapsed time since the formation of the first protostar.
Next, we compare this trend with a number of published numerical studies by scaling the elapsed time according to the scale-free nature of the system. As a result, we find most of the results in the literature agree well with the relation.
Present results combined with the previous studies in the literature imply that the PopIII stars tend to be born not as single stars, but in multiple systems. 
\end{abstract}

\section{Introduction}
Formation theory of the first stars predicts that very massive star formation of mass $M \gtrsim 100M_\odot$ is preferred in primordial environments \citep{omukai98}.
In the $\Lambda$CDM paradigm, those stars form in minihalos of $M\sim 10^6M_\odot $ at $z\sim 20-40$~\citep [e.g.,][]{haiman96,nishi_susa99,fuller_couchman00,abel02, bromm02, yoshida03}. Pristine gas in the minihalos cool via H$_2$ line cooling, which is significantly less efficient than cooling by heavier elements and dust. As a result, the gas temperature is higher than the present-day counterpart by two orders of magnitude. For the gas cloud to 	collapse, the mass of the cloud should exceed the Jeans mass, which is larger for higher temperatures. Consequently, the mass of the collapsing primordial gas is of the order of $10^3 M_\odot$ at $\sim 10^4~{\rm cm^{-3}}$, while it is $\sim 1M_\odot$ in the local environment. Therefore, it seems to be reasonable that very massive stars are preferred in the primeval environment as a first approximation.

On the other hand, recent numerical simulations of the mass accretion phase predict that the disk around the primary protostar fragments into multiple pieces that could culminate in binaries, triplets, etc.~Thus, the number of these fragments per minihalo increases, and the mass distribution of them also changes dramatically\citep{stacy10,stacy12,stacy16,clark11a,greif11,greif12,smith11,hirano17}.  These fragments, the population III (PopIII) protostars, could be less massive than $100M_\odot$, or even less than 0.8 $M_\odot$, which could survive until the present. To test for the presence of such low-mass stars, it is useful to search for the zero-metallicity stars in our galaxy \citep{hartwig15a,ishiyama16,griffen18,magg18,magg19}. It is also important to match the theoretical initial mass function (IMF) to the abundance ratio observed on the extremely metal poor (EMP) stars \citep[e.g.][]{susa14,ishigaki18}, because they could be born in the remnants of first stars \citep[e.g.,][]{smith15,chen17,chiaki18}. Such efforts can constrain the mass distribution of the first-generation stars.
In any case, it is crucial to have a good theoretical assessment of the mass distribution  of the PopIII stars  to compare with observations.
 
However, the number of the protostars per minihalo and the mass distribution of them are still unclear from numerical studies. The final mass of the fragment depends fundamentally on the radiative feedback by the protostars themselves, which requires long integration times.
The feedback arises after the protostars have grown to $15-20M_\odot$, which corresponds to several thousand years after the formation of the first protostar. After the onset of the radiative feedback, the accreting gas is heated by the UV radiation and evaporated. Finally, the gas accretion onto the protostars is shut off to yield the final mass of the protostars.
This result can only be achieved by coarse radiation-hydrodynamics simulations, which cannot trace the detailed fragmentation and merger processes at the beginning of protostar formation \citep{hosokawa11,stacy12,susa13,susa14,hirano14,hirano15,stacy16,hosokawa16}. 

Alternatively, high resolution simulations that resolve the initial phase of the protostar formation, reveal that the fragmentation of the accretion disk as well as the merging of the fragments are quite common, though only for about 10-100 yrs after primary protostar formation can be traced because of the high numerical cost.
For instance, the highest resolution cosmological simulation by \citet{greif12}, without sink particles, has shown that $\sim$1/3 of the formed fragments survive until $\sim$10 yrs after the formation of the primary protostar, while another $\sim$2/3 merge with each other.  
They predict that about four fragments survive on average through the simulated time.
\citet{machida_doi13} also have performed similar calculations to find $\sim 10$ fragments, and many of them merge in their nested-grid highest resolution box, but some are ejected to outer low-resolution regions where their fate cannot be traced.

Sink particle simulations also predict fragmentation of the disk, and they follow relatively longer time scales than non-sink simulations. For example, \citet{stacy16} trace the evolution of a single minihalo for 5000 yrs, and they also include the effects of ultraviolet radiative feedback by the forming protostars, in the later phase of the evolution. They predict $37$ sinks, that is, protostars, are expected in a minihalo. However, the differences in numerical methods, i.e., the calculations with or without sink particles, might affect the final results.

In addition, all non-sink/sink simulations introduce an artificial threshold density above which the gas clump is replaced by a sink particle or the equation of state becomes stiff to avoid the very short time scale inside the protostars\footnote{One exception is the calculation by \citet{greif12}. They solve the radiative transfer of cooling photons, thereby introducing the threshold density naturally.}. This threshold density is very different among studies, depending on the time scale that each author tries to resolve. As a result, the results are so diverse that it is difficult to derive a universal law of fragmentation of PopIII accretion disks.

In this study, we perform another simple hydrodynamics simulation with and without sink particles to investigate the formation and merger of the fragments in the accretion disk surrounding protostars in a primordial environment. 
We analyze the numerical results in detail, and compare them with a number of published results by using a scaling relation. Utilizing all the numerical results so far, we infer the final number of fragments in the minihalos.

\section{Methodology}
\subsection{Simulations without sinks}
We employ the standard SPH scheme to calculate the dynamics of a collapsing gas cloud. 
The code used in this study is based on RSPH \citep{susa04,susa06}, designed to solve the chemical network as well as the radiative cooling/heating/transfer of primordial gas. In this study, the chemistry/cooling/transfer solvers are switched off. Instead, we simply assume the barotropic relation between the density and the temperature of the gas cloud to save computational time. The barotropic equation of state is pre-calculated by a one-zone model \citep{susa15} and tabulated to be used in the hydrodynamics simulations.

Because we are interested in the mass-accretion phase, we have to set a threshold hydrogen number density, $n_{\rm th}$, above which the gas becomes stiff against further collapse. In the present study, we set the threshold density at $n_{\rm th} = 10^{15}~{\rm cm^{-3}}$, above which the relation $T \propto n_{\rm H}^{\gamma_{\rm eff}-1}$ holds, where $\gamma_{\rm eff}=5$. 
This very stiff equation of state is employed in opposition to the sink particles simulations, which introduce a ``hole'' around the sinks.

The initial condition of the simulation is a {Bonnor}-Ebert sphere around the critical density of H$_2$ level transition, which mimics the ``loitering'' primordial gas cloud found in cosmological simulations \citep{abel02,yoshida03}. The particle distribution is created as follows. First, we put particles in a box to relax until the system settles down to a uniform and equilibrium state. As a result, we have a glass-like nearly uniform particle distribution instead of an aligned distribution on grid points. The density fluctuation due to the glass-like distribution is $\lesssim 3\%$. Then, we hollow out a spherical region. Finally, the radial distribution of the particles is transformed to generate the isothermal {Bonnor}-Ebert sphere of $T=200$ K at $n_{\rm H}=10^4 {\rm cm^{-3}}$.  To boost the collapse, we enhance the mass by 40\% when the simulation is started. The resultant total mass of the system is $1.88\times 10^3 M_\odot$. As a result, the initial central density of the cloud is $1.4\times 10^4{\rm cm^{-3}}$, and the temperature is 196K assuming barotropic relation.
We also add a rigid body rotation to the sphere of angular velocity $\Omega_{\rm rot}=2\times 10^{-14}~{\rm s^{-1}}$, 
that corresponds to the ratio of rotational energy to gravitational energy of $T/|W|\simeq 0.05$.
As a result, the specific angular momentum at the outer edge of the cloud ($\sim 2$pc) is $1.2{\rm pc\cdot km/s}$, which is consistent with a typical case of collapsing primordial gas in minihalos \citep{yoshida06}. 
At this point, the cloud is spherical, thereby we can freely choose the direction of the angular momentum vector. We choose five directions randomly, to represent slightly different realizations due to the particle distribution. These five realizations are labeled as R1$\sim$R5. These five almost identical runs with slightly different realizations can provide the knowledge how chaotic nature of the system can affect the final outcome of the numerical calculation.
\begin{table}
\caption{Summary of numerical parameters}
\begin{center}
\begin{tabular}{cc}
\hline
cloud mass & $1.88\times10^3 M_\odot$\\
initial angular velocity & $2\times 10^{-14}{\rm s^{-1}}$\\
rotation axis& randomly chosen for R1-R5 runs\\
initial central density & $1.4\times 10^{4} {\rm cm}^{-3}$\\
initial central temperature & 196 K\\
$m=2$ perturbation & $\delta = 0.1$\\
\hline
maximal mass resolution &$0.014M_\odot$\\
$n_{\rm th}$ & $10^{15}{\rm cm^{-3}}$\\
\hline
\end{tabular}
\end{center}
\tablecomments{Numerial parameters for R1-R5 and an SPH run. SPH run uses identical initial condition of R1.}
\label{tab:runs}
\end{table}

We also add anisotropy in particle position, that is, we transform the SPH particle position:
\begin{eqnarray}
x&\rightarrow& x\left(1+\delta \cos2\phi\right)\\
y&\rightarrow& y\left(1+\delta \cos2\phi\right) ,
\end{eqnarray}
where $x$ and $y$ denote the Cartesian coordinates of SPH particles perpendicular to the rotation axis, $\phi$ denotes the azimuthal angle in the $xy$ plane, and $\delta =0.1$. This perturbation introduces a slightly flattened particle distribution in the $xy$ plane of mode $m=2$.
The initial mass of each SPH particle is $1.8\times 10^{-3} M_\odot$, and it is split into 13 particles \citep{kitsionas02} when the central density exceeds $10^{10}~{\rm cm}^{-3}$, in case the particles are within the distance of $5\times 10^{-2}$ pc from the center. Because the number of neighbor particles in the SPH scheme is chosen as 50, the mass resolution at the central part of the cloud after the split is $2N_{\rm neib}m_{\rm SPH}= 0.014 M_\odot$ \citep{bate_burkert97}, where $N_{\rm neib}$ and $m_{\rm SPH}$ denote the number of neighbor particles and the mass of an SPH particle, respectively. Basic parameters are summaried in Table\ref{tab:runs}.

When the central density reaches $n_{\rm th}$, we cut out the central spherical region with $r_{\rm cut}=0.015$ pc to follow the subsequent evolution.
Larger $r_{\rm cut}$ enables us to follow longer time evolution at a cost of longer computational time. 
We can follow the evolution until $\sim 1600$ yr after the central core forms for $r_{\rm cut}=0.015$ pc, after which the absence of an outer envelope reduces the mass accretion rate onto the central region. 
\subsection{Simulations with sinks}
We also perform a sink particle simulation for reference. We perform a run corresponding the same initial conditions of R1.
The sink prescription is a standard one. First, we pick up SPH particles where the density exceeds $n_{\rm th}$ and the SPH smoothing length is less than the accretion radius $r_{\rm acc}$. Here, we set $r_{\rm acc}=3$ AU, which corresponds to the Jeans length at $n_{\rm th}$, assuming the barotropic relation. Next, we identify the particles that correspond to the potential local minimum, i.e., find the particles that have lowest gravitational potential among the neighboring particles. 
Finally, we check that the total energy of the particle and its neighbor particles as a clump is negative. The kinetic energy of the motion of the center of mass is not included in the energy budget. SPH particles that pass all these tests are changed into sink particles. 
After the sink formation, if SPH particles are closer to a sink particle than the accretion radius $r_{\rm acc}$ and if they are gravitationally bound by the sink in a two-body relationship, they fall onto the sink particle to add their mass and linear momentum to the sinks. Sink--sink mergers are also treated in the same manner as the SPH-sink merger procedure.

\section{Results}
\subsection{Run-away phase}
\begin{figure}[htbp]
	\centering
	{\includegraphics[width=8cm]{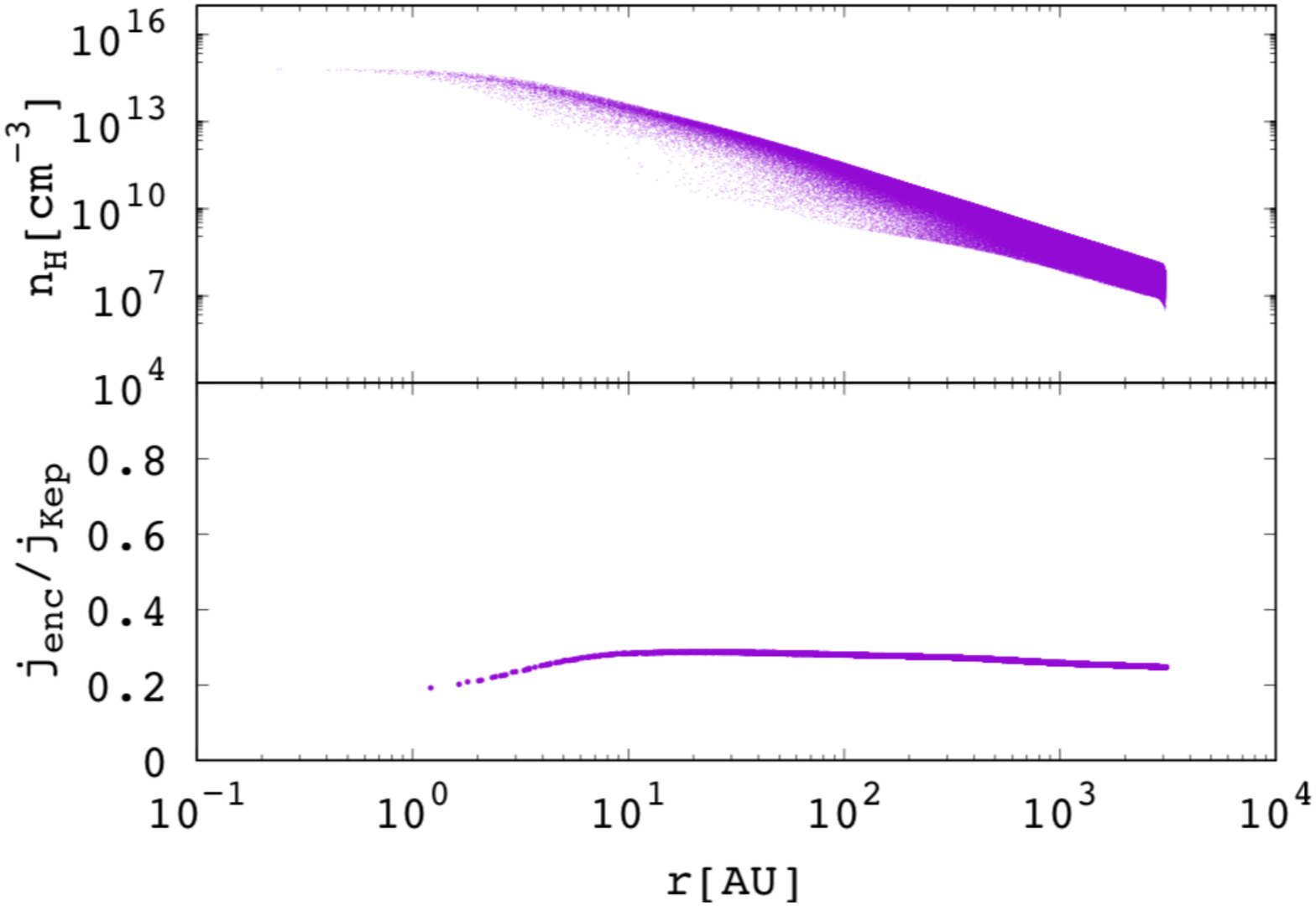}}
	\caption{Upper panel: Density distributions for run R1 at the end of the runaway phase. Horizontal axis is the distance from the density peak. Lower panel: Radial distribution of the ratio of the enclosed specific angular momentum and the Kepler value defined as $\sqrt{GM(r)r}$, where $M(r)$ denotes the mass enclosed within the radius $r$.}
	\label{fig:runaway}
\end{figure}

Fig.\ref{fig:runaway} is an image at the end of the runaway phase. The upper panel shows the density distribution of the run R1. The direction of the spin vector does not affect these results in the runaway collapse phase, therefore we do not plot the results of runs R2$\sim$R5. We find a typical distribution for a similarity solution with $\gamma_{\rm eff} \simeq 1.09$, that leads to the envelope density distribution of $\propto r^{-2.2}$ \citep[e.g.][]{suto_silk}. This is a well-known behavior of collapsing primordial gas clouds in cosmological simulations \citep[e.g.,][]{yoshida06}. The lower panel shows the specific angular momentum distribution. The plotted specific angular momentum values are averaged within a sphere of a given radius $r$, and are normalized by the Kepler rotation value, $j_{\rm Kep}=\sqrt{GM(r)r}$,  where $M(r)$ denotes the mass enclosed within the radius $r$. We find this ratio is roughly constant at $\sim 0.3$ in the envelope. These are again consistent with similarity solutions \citep{saigo98}. 
\begin{figure*}[bth]
	\centering
	{\includegraphics[width=16cm]{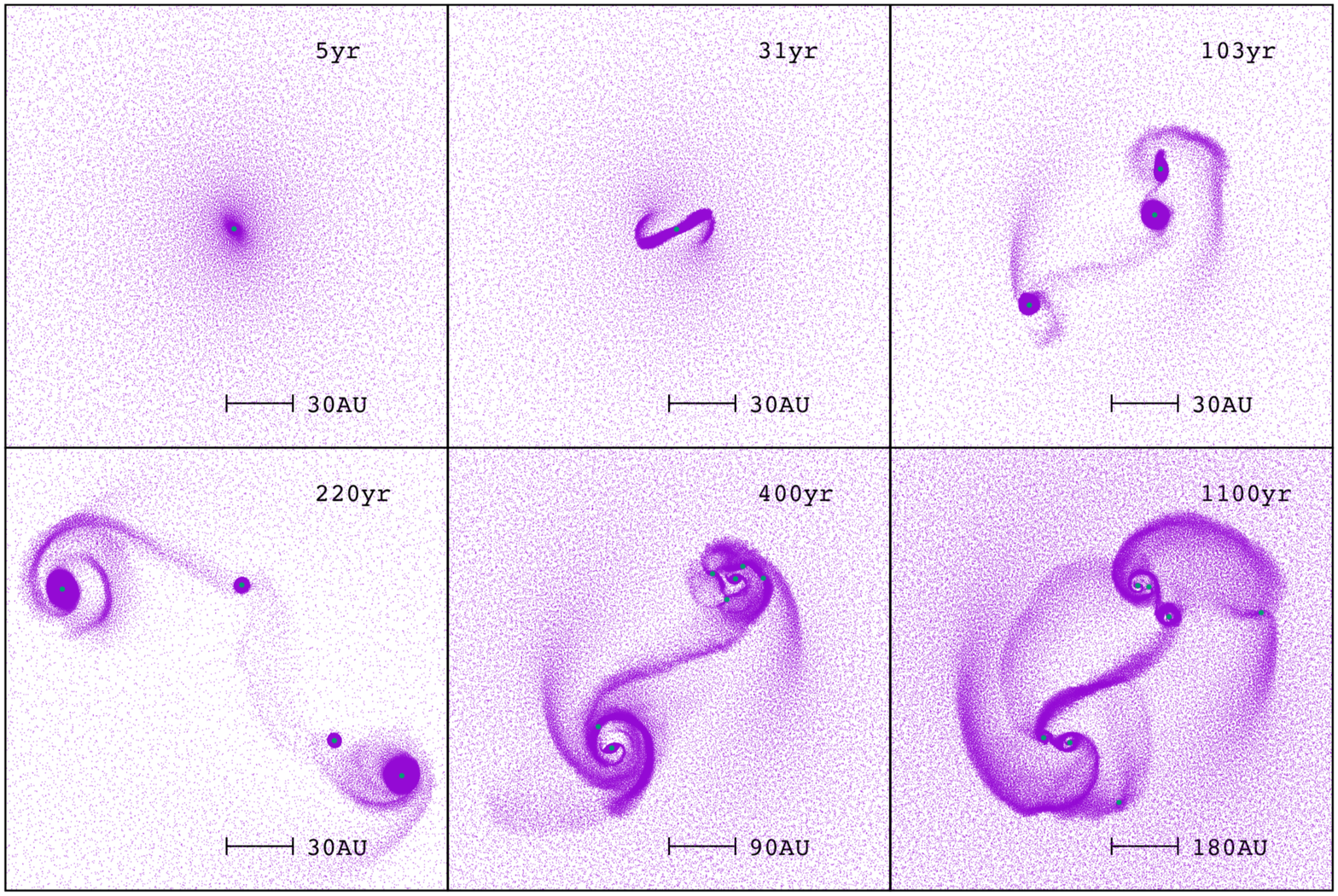}}
	\caption{Images of the face-on view of run R1. The corresponding times are (left-to-right, top-to-bottom) $t=5, 31, 103, 220, 400, {\rm and } 1100$ yrs. Green dots denote the position of the identified gas clumps.}
	\label{fig:fragmentation}
\end{figure*}
\subsection{Disk fragmentation in the mass accretion phase}
\label{fragmentation}
After the central number density reaches $n_{\rm th}=10^{15}{\rm cm^{-3}}$, collapse of the central region is stopped by the enhanced pressure owing to the stiff equation of state. Thus, the evolution of the cloud shifts into the mass accretion phase. As an overview of the accretion phase, Fig.\ref{fig:fragmentation} shows the time evolution of run R1. The six panels are face-on views of the formed disks, corresponding to times $t=5, 31, 103, 220, 400, 1100$ yrs, respectively. The origin of this time coordinate is the epoch at which the central density exceeds $n_{\rm th}$ for the first time. The purple dots represent the positions of SPH particles, and the green dots denote the positions of identified fragments. The identification procedure is as follows. First, we identify SPH particles of number density higher than $0.9n_{\rm th}$. Then the SPH particles which are neighbor particles are connected with each other, to create ``islands'' of dense regions. Finally, if the number of SPH particles in a clump exceeds $6N_{\rm neib}$, the clump is identified as a fragment, shown by the green dots. This number corresponds to a mass scale slightly higher than three times the Jeans mass at $n_{\rm th}$. Hence, the identified fragments are not temporarily dense clumps but are gravitationally bound objects. 

First, the central dense region reaches the threshold density $n_{\rm th}$, a dense clump forms (5yr). After a while, the accreted gas forms two spiral arms and the system is still symmetric (31yr). In this phase, the elongated structure is identified as a single clump. This structure arises because of the $m=2$ mode fluctuation added initially. Then the arms fragment into two fragments so we have three fragments(103 yr). We also see that the symmetry is lost at this stage, and the primary fragment starts departing from the center. This asymmetry is due to the tiny density fluctuation due to the glass-like distribution of the SPH particles. Since the rotation axies of runs R1-R5 are different with each other, so different seed density perturbations are embedded in the runs.

Throughout the evolution of the system, we find vigorous fragmentations of the formed disk as well as the merging of many fragments.
 We also see that the size of the disks-plus-fragments system is growing with time (220yr-1100yr). This is because the accreting matter brings in the orbital angular momentum from the envelope. The fragmentation proceeds mainly in the individual accretion disk around each fragment, as discussed in the next subsection.
\begin{figure}
	\centering
	{\includegraphics[width=10cm]{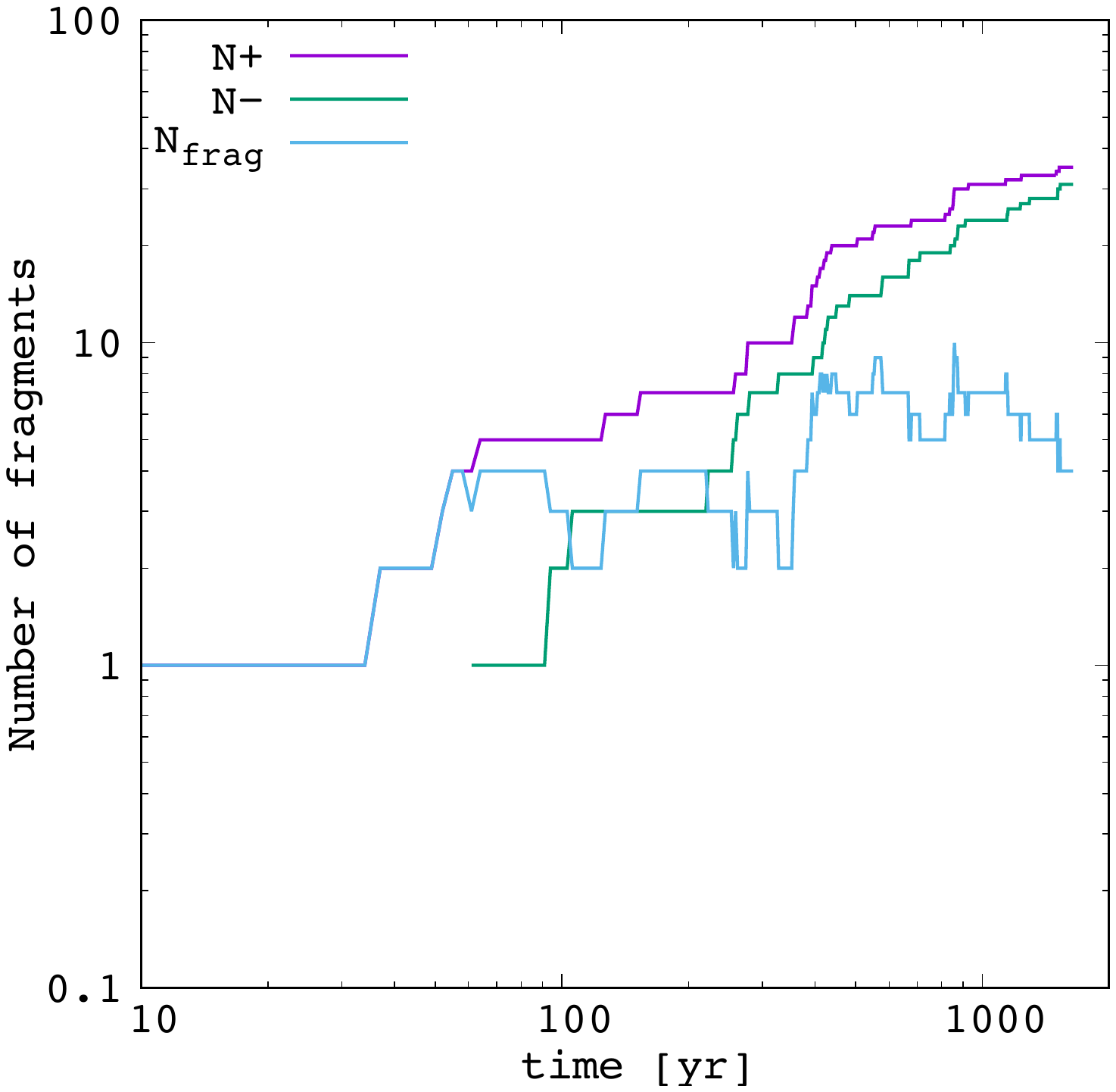}}
	\caption{Number of fragments ($N_{\rm frag}$), cumulative number of formed fragments ($N_+$) and cumulative number of merger events ($N_-$) as  functions of time in run R1.}
	\label{fig:dndt}
\end{figure}
\begin{figure}
	\centering
	{\includegraphics[width=8cm]{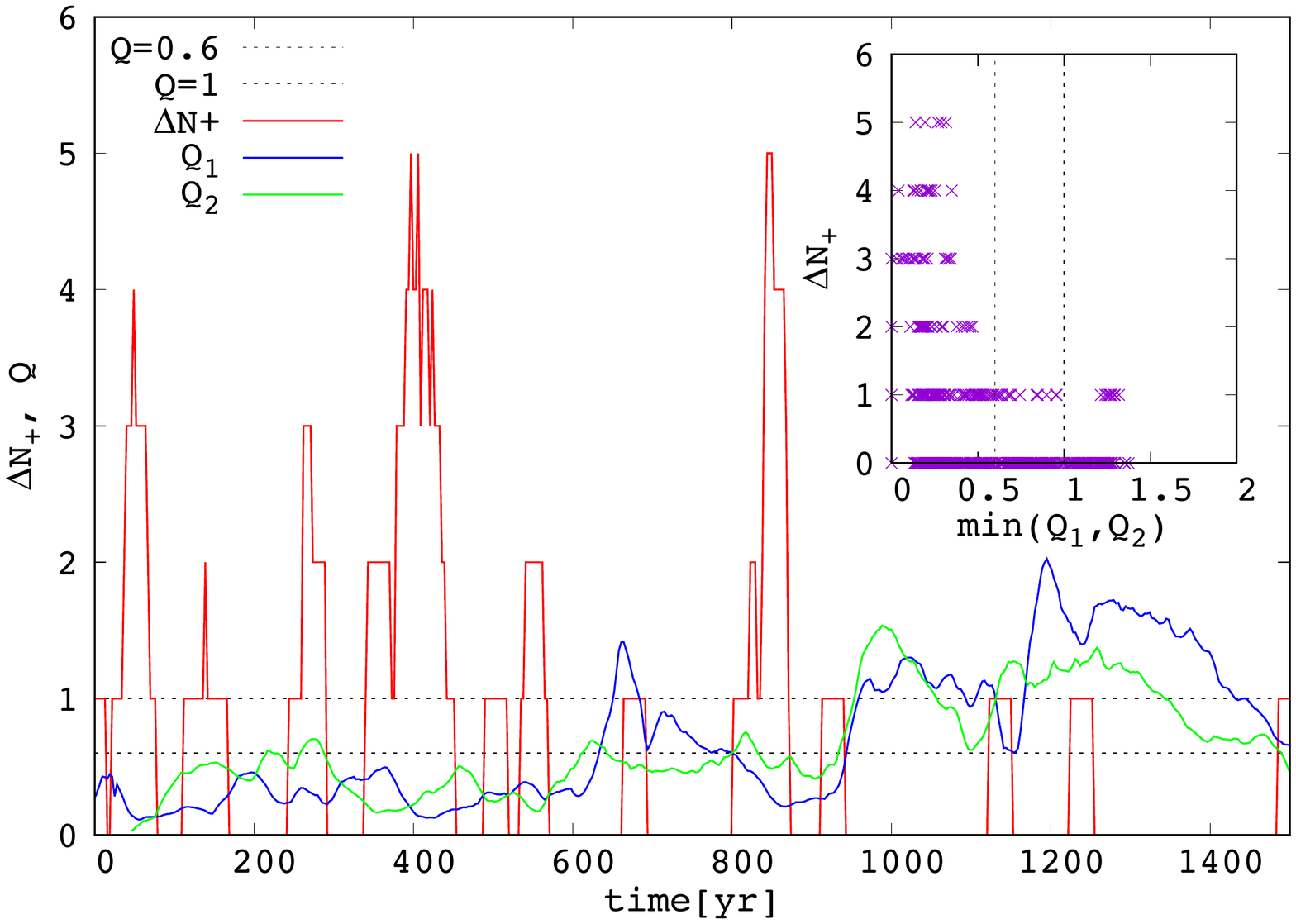}}
	\caption{Time evolution of the minimal Toomre Q parameters in the disks around the most massive two fragments (blue and green curves) in  run R1. Red curve shows the number of formed fragments within a short interval of $\Delta t= 30$ yrs.}
	\label{fig:toomre}
\end{figure}
Fig.\ref{fig:dndt} shows the evolution of the cumulative number of fragmentation events ($N_+$) and mergers ($N_-$), along with the number of fragments ($N_{\rm frag}$) for run R1.  
 We find the relation $N_+ \gtrsim N_-$ holds during the evolution, and the slight difference causes a gradual increase of the number of fragments. 
 {In Fig.\ref{fig:toomre}, the red curve shows the time evolution of the number of formed fragments ($\Delta N_+$) within a short interval of $\Delta t= 30$ yrs.  The blue and the green curves show the minimal Toomre Q parameters $Q_1$ and $Q_2$ in the disks around the most massive two fragments in run R1. The Q-values are also smoothed over $\Delta t= 30$ yrs.} The fragmentation process is episodic, and the fragmentation is active when $Q \lesssim 0.6$. {In the superimposed panel, $\Delta N_+$ at every snapshot is plotted against the minimal of $Q_1$ and $Q_2$. We find that in case the fragmentation is active ($\Delta N_+ \gtrsim 1$), the Q value is tend to be less than $\sim 0.6$.} This condition, $Q < 0.6$, which is derived by \citet{takahashi16}, corresponds to the condition of gravitational instability of the spiral arms in the disk, while $Q\lesssim 1$ denotes the condition of spiral arm formation. Hence, the present fragmentation process is due to the gravitational instability of the accretion disks.

\begin{figure}
	\centering
	{\includegraphics[width=10cm]{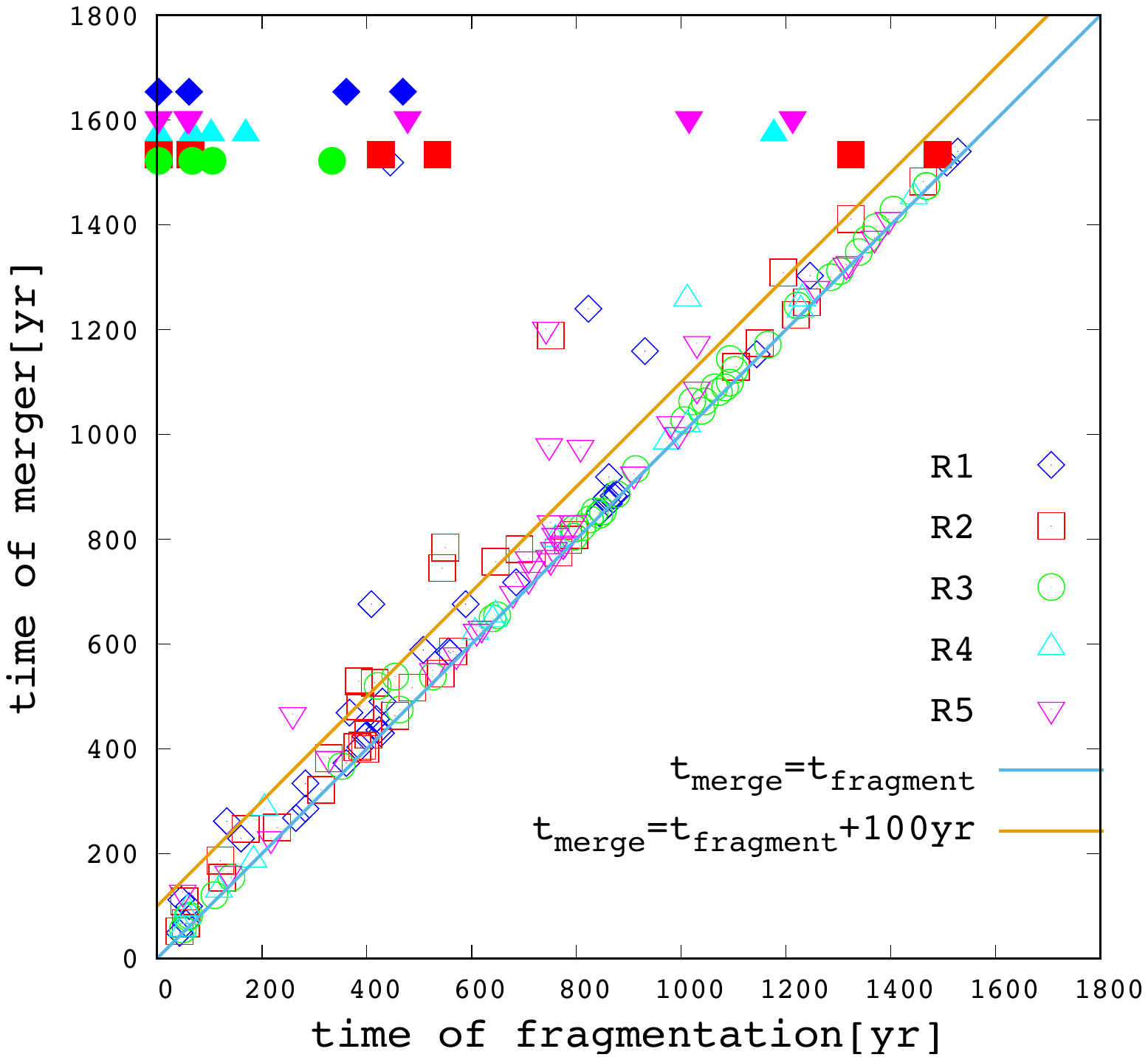}}
	\caption{Time of fragmentation vs. time of merger for all fragments in runs R1-R5. Filled symbols show the survived cases, while the open symbols show the merged ones. For the survived cases, the final time of simulation is used instead of the time of merger.}
	\label{fig:survive}
\end{figure}

Fig.\ref{fig:survive} shows the time of fragmentation vs. the time of merger for all fragments in the runs R1-R5. Because the fragments merge after their birth, all points are located in the upper left region. It is evident that most of the fragments are just above the diagonal line, where the time of merger almost equals the time of fragmentation. This means the survival times of most fragments are very short compared to the time for the fragmentation. The typical short survival time is roughly $\lesssim 100$ yrs, which is comparable to the typical orbital period of each accretion disk, and is consistent with previous results \citep[e.g.][]{greif12,hosokawa16,hirano17}. Such a short survival time and the episodic nature of disk fragmentation by gravitational instability leads to the picture discussed by \cite{vorobyov13}, in which the fragmentation occurs when the disk mass is sufficiently loaded, followed by the rapid migration process.
{However, we also find 4-6 fragments in each run that survive until the end of the simulations ($\sim 1600$ yrs), shown by the filled marks.  Hence, most of the fragments merge at a very short time scale, but some fraction survives. The survived fragments are rare among all of the fragmentation events but exist in all the runs (R1$\sim$R5).}

\begin{figure}
	\centering
	{\includegraphics[width=10cm]{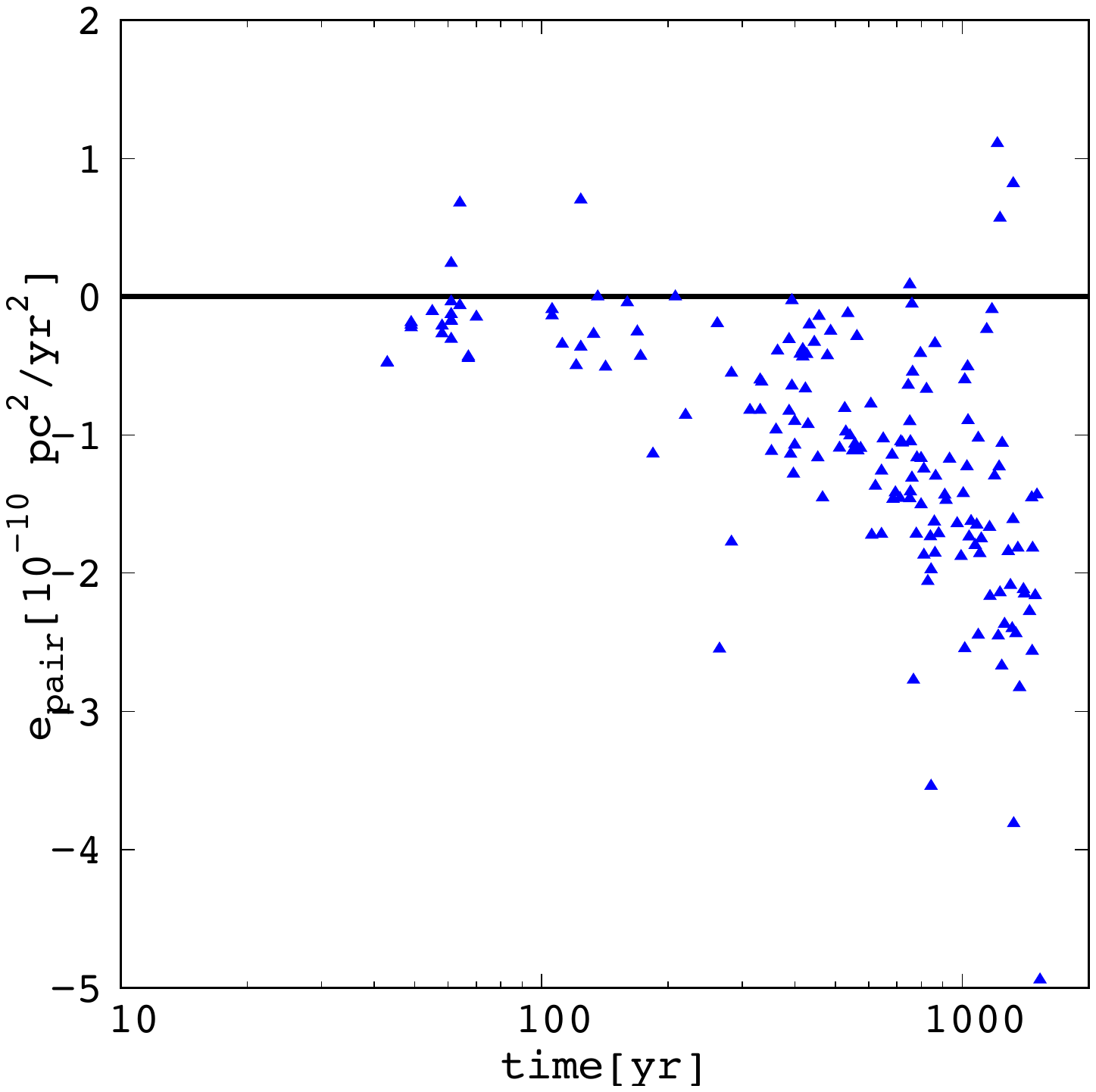}}
	\caption{Time vs. $e_{\rm pair}$ for all fragments in the runs R1$\sim$R5.}
	\label{fig:binding_energy}
\end{figure}
\begin{figure}
	\centering
	{\includegraphics[width=10cm]{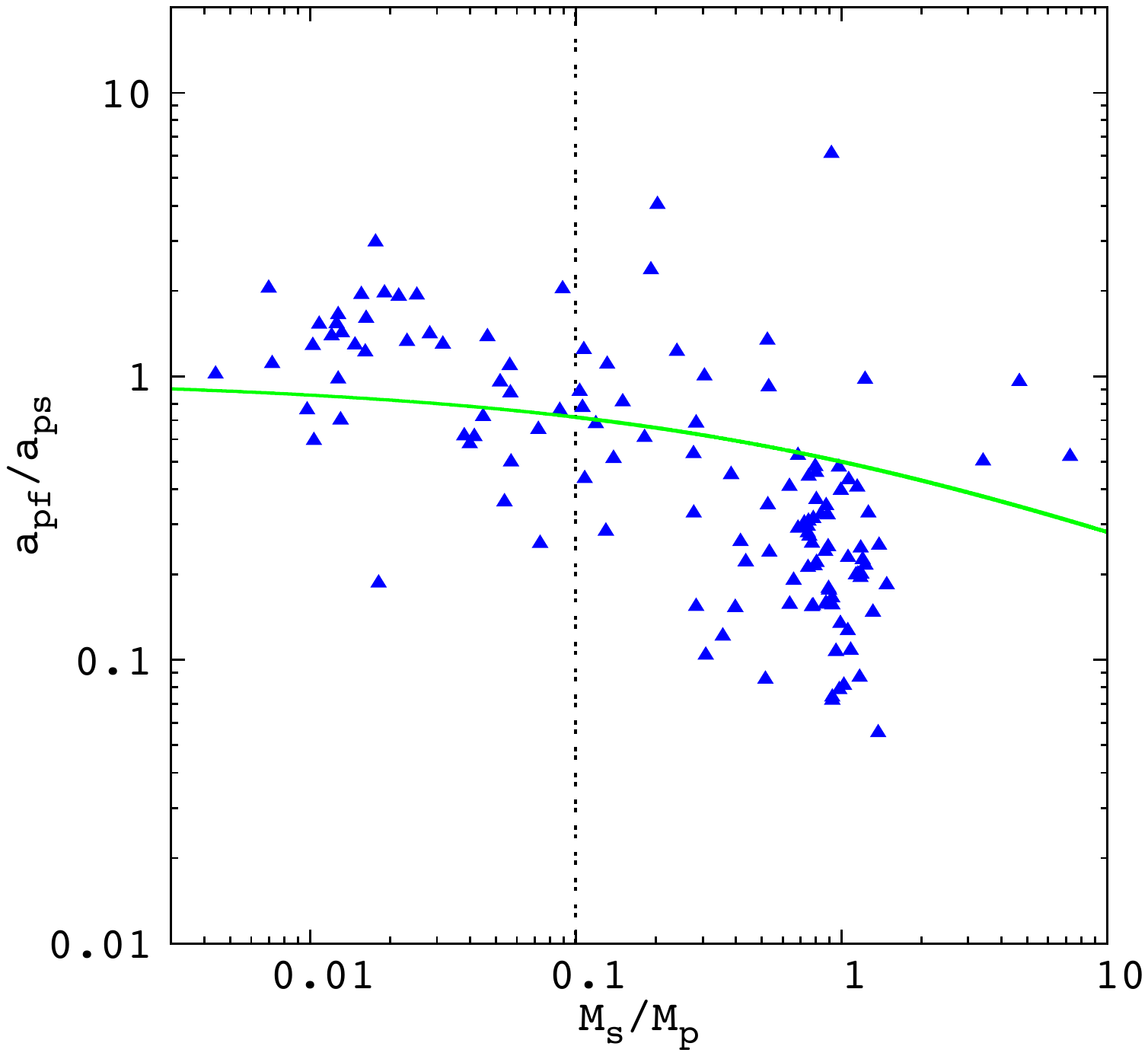}}
	\caption{Fragments born in binary systems are plotted on $a_{\rm pf}/a_{\rm ps}$ vs. $M_s/M_p$ plane. Green curve shows the distance from the primary star and the L1 point in the binary system. Vertical dashed line $M_{\rm s}/M_{\rm p}=0.1$ roughly indicates the ``planet like'' cases.}
	\label{fig:frag_location}
\end{figure}

\subsection{Position of fragmentation}
To have a better understanding of the fragmentation process, we assess 
the two-body specific binding energy $e_{\rm pair}$ for each fragment at its birth with the pre-existing fragments. 

The specific two-body binding energy $e$ is described as:
\begin{equation}
e= \frac{1}{2}\left(\bm{v}_{\rm frag} -\bm{v}_{\rm pre}\right)^2 -\frac{G \left(m_{\rm frag}+m_{\rm pre}\right)}{|\bm{r}_{\rm frag} -\bm{r}_{\rm pre}|}
\end{equation}
where $\bm{v}, \bm{r},$ and $m$ denote velocity, position and mass, respectively, whereas the subscripts ``frag'' and ``pre'' denote the newly formed fragment and the pre-existing one. {We identify all the pre-existing fragments those who have negative $e$ with a newly formed fragment.}  We define $e_{\rm pair}$ of a newly formed fragment as the two-body binding energy with the nearest pre-existing fragment among the fragments with negative $e$. In case we find no pre-existing fragment that has negative $e$ with the newly formed fragment, we just use $e$ with the nearest pre-existing fragment. 
Fig.\ref{fig:binding_energy} shows $e_{\rm pair}$ for all the newly formed fragments in all runs R1$\sim$R5. The horizontal axis shows the time of fragmentation.
It is quite clear most of the fragments are bound at their birth by one of the fragments that formed earlier. In fact, 96\% of the fragments have negative $e_{\rm pair}$ at their birth.
In other words, most of the fragments are born as a companion of another fragment. 

To examine the environment of the fragments at their birth more closely, we check whether the pre-existing companions have another companion prior to the newly formed fragment under consideration.
In fact, they do have {previous} companions in 81\% of all the fragmentation events, while 15\% of the fragmentation events occur around a single pre-existing fragment. For the 81\% events, we plot the distance from the newly born fragments to their companion (we call this a ``primary'' star), normalized by the distance between the primary star and its pre-existing companion star (hereafter we call this a ``secondary''. Note that the terminologies of primary and secondary here are different from the usual definition).
In Fig.\ref{fig:frag_location}, the horizontal axis denotes the mass ratio of the primary and secondary. Note that this ratio can be larger than unity by its definition. For small mass ratio, i.e., in the left side of the figure, the pre-existing secondary is far less massive than the primary. Hence, it does not affect the motion of the newly formed fragment around the primary. This means that the newly formed fragment behaves like a ``planet,'' as does the secondary. In contrast, in the right side of the figure, the mass of the primary is comparable to or less than that of  the secondary. In this case, most of the fragments are born in the neighborhood of the primary. The solid curve denotes the distance from the primary to their shared Lagrangian L1 point defined for the primary and the secondary. The majority of the fragments are located within the L1 radius, that is, they are within the gravisphere of their primaries. Hence, they are born in the accretion disk of the primary. In other words, most of the fragmentation events occur in the accretion disk associated with each fragment, but not in the circumbinary disk nor ``circum-multiple'' disk.
\begin{figure}
	\centering
	{\includegraphics[width=10cm]{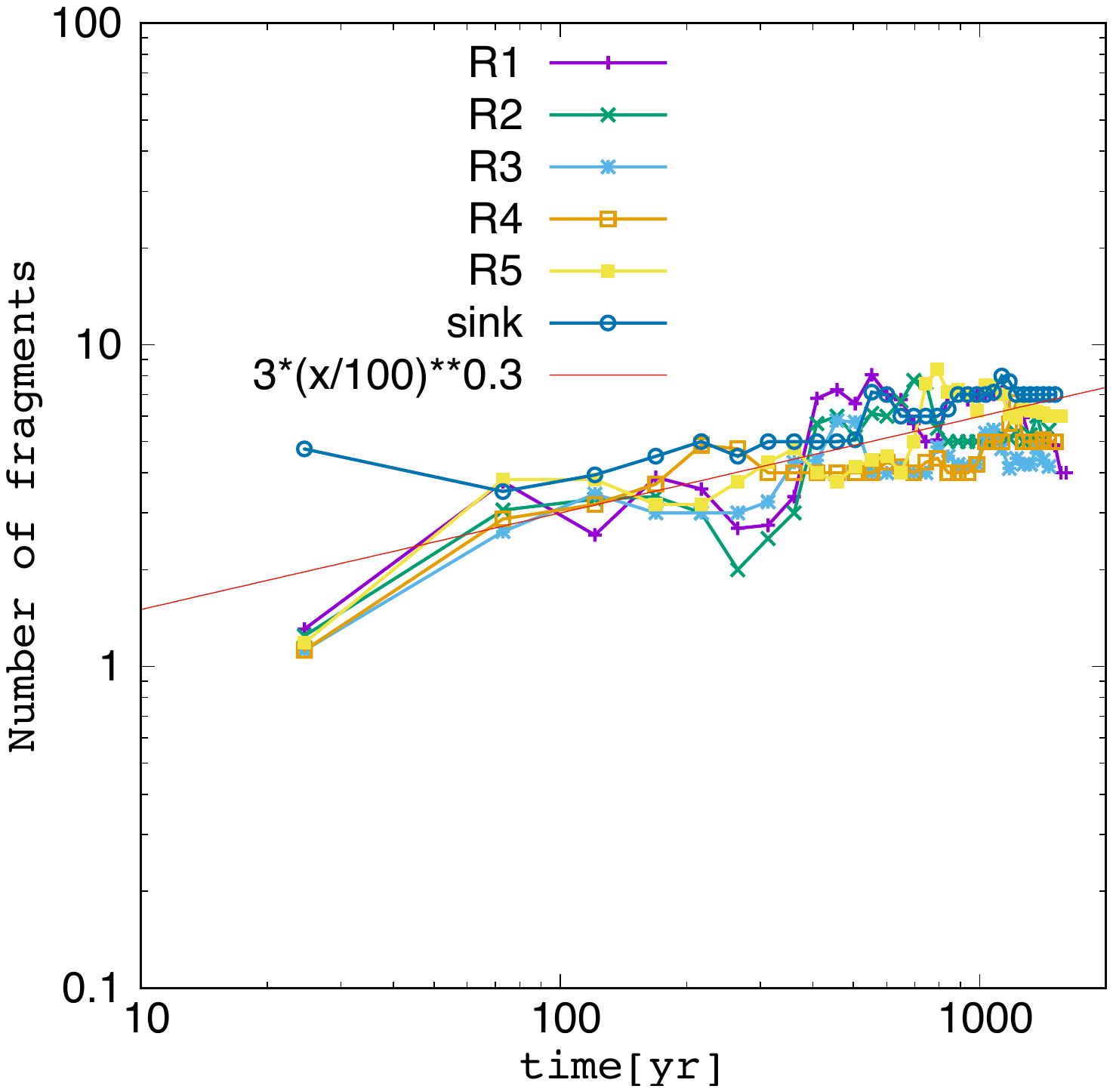}}
	\caption{Number of fragments in the non-sink runs (R1$\sim$R5) and a sink run (R1)}
	\label{fig:nfrag}
\end{figure}
\begin{figure}
	\centering
	{\includegraphics[width=10cm]{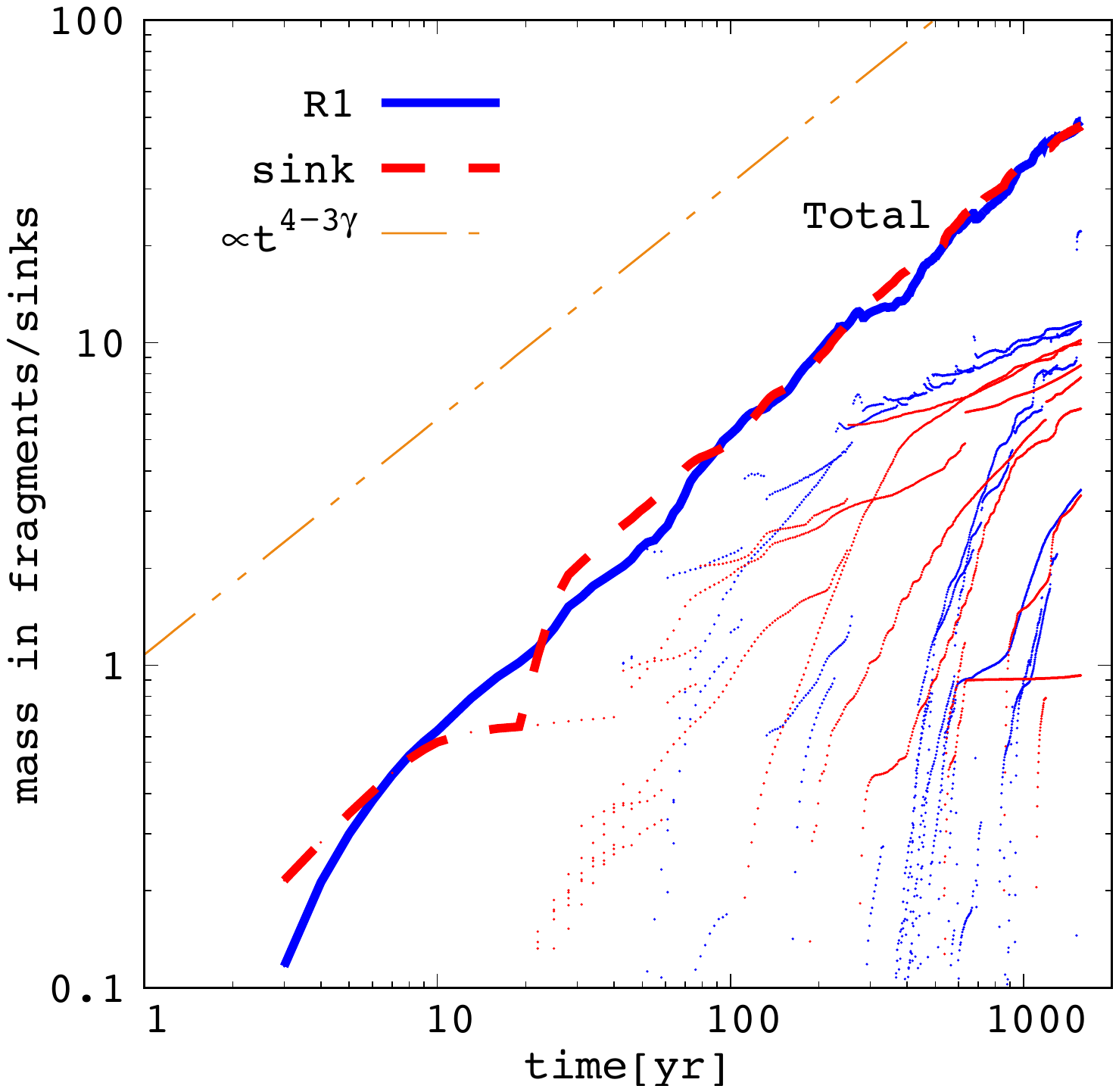}}
	\caption{Total mass accreted onto the identified fragments and sink particles as functions of time {are shown by thick curves. Mass of the all fragments/sinks are also shown by dots.}}
	\label{fig:mass_tot}
\end{figure}

\subsection{Number of fragments}
In this section, we attempt to model the evolution of $N_{\rm frag}$ by a simple equation and compare it with numerical results.
As we discussed in section \ref{fragmentation}, fragmentation occurs when the $Q$-value of each accretion disk becomes low enough, which implies the fragmentation proceeds in a mass-loading time scale of the disks \citep{vorobyov13}. The mass-accretion time scale of the whole system from the envelope is $M(r)/\dot{M}$, where $M(r)$ denotes the enclosed mass within a radius $r$, and $\dot{M}$ is the total mass accretion rate of the system. This time scale is nearly proportional to $t$, which is the elapsed time since the first protostar formation. Therefore, we simply assume the fragmentation time scale is proportional to $t$.  
Because most of the fragmentation occurs in the accretion disk around each protostar, the formation rate of the fragments is proportional to the number of fragments, $N_{\rm frag}$, and as a  result, we assume the formation rate of the fragments is
\begin{equation}
\left(\frac{dN_{\rm frag}}{dt}\right)^+ \propto \frac{N_{\rm frag}}{t}.
\end{equation}
It is clear that the merger process proceeds very rapidly (Fig.\ref{fig:survive}), and the merger events quickly follow the fragmentation. Here, we simply assume that the ``survival fraction'', which is the fraction of the newly formed fragments to survive the merger processes, is constant.
Consequently, the time evolution of $N_{\rm frag}$ is described as
\begin{equation}
\frac{dN_{\rm frag}}{dt} = p
\frac{N_{\rm frag}}{t} , \label{eq:basiceq}
\end{equation}
where $p$ is a non-dimensional constant. This equation can be integrated easily, yielding
\begin{equation}
N_{\rm frag} \propto t^p . \label{eq:basicsol}
\end{equation}
Thus, if we accept this simple equation (\ref{eq:basiceq}), the number of fragments evolves in a power law solution as a function of time after the formation of the first protostar.
In Fig.\ref{fig:nfrag},
the number of fragments $N_{\rm frag}$ for all non-sink runs { (R1$\sim$R5)} are shown, as well as the results from the run with sinks. For the sink particle simulation the number of sink particles is regarded as $N_{\rm frag}$. Here, the number is averaged over time intervals for every {50} yrs. In all cases, $N_{\rm frag}$ grows slowly as a function of time, with a scatter due to the {continual} fragmentation and merger processes. It also should be noted that the differences due to the slightly different realizations {(R1-R5)} cause some scattering in the results, but it is not large.  If we fit the results with a simple power of $t$, we have $N_{\rm frag}\propto t^{0.3}$, which is consistent with the analytical model.

\subsection{Comparison of runs with stiff-EOS and that with sinks}
{As mentioned in the previous subsection, the results of the sink simulation is also plotted in Fig.\ref{fig:nfrag}. The time evolution of the number of sinks/fragments are similar with each other, except in the initial phase of $t \lesssim 100$yrs. In the sink simulation, we find a ring just outside the accretion radius, that becomes gravitationally unstable to form several sinks eventually. This would be an artifact of the present sink procedure. This could be avoided by suppressing the sink formation within two times the sink radius \citep[e.g.][]{clark11a}. However, these sinks merge quickly with each other to settle down to the track in Fig.\ref{fig:nfrag} consistent with the results from stiff-EOS runs. In addition, we do not observe this kind of artifact in the later phase of the evolution.

Fig.\ref{fig:mass_tot} shows the time evolution of the total mass accreted onto sink particles or fragments identified in the non-sink simulation R1 {(thick curves)}. The results from two runs are fundamentally same, although there are some deviations with each other. The total mass increase as a power of the elapsed time. The mass of the central singularity in the spherical similarity solution of $\gamma_{\rm eff}$ is poroportional to $t^{4-3\gamma_{\rm eff}}$, which is also shown in the figure as a guide for the eye assuming $\gamma_{\rm eff} =1.09$.

Fig.\ref{fig:mass_tot} also shows the mass of all fragments/sinks in each simulation. 
The basic behaviour of the two runs are same, although the individual evolution is very different. The difference comes from the initial ring fragmentation in the sink simulation that is not found in the stiff-EOS runs and also is due to the chaotic nature of the system. In fact, the number evolution is not identical to each other even in stiff-EOS runs R1-R5(Fig.\ref{fig:nfrag}).

The final number of this particular two runs are 4 (stiff-EOS) and 7 (sink). On the other hand, the final number of fragments in the stiff-EOS runs (R1-R5)  spread over 4 to 6.
Hence 7 sink particles seems to be upward, but not so significant.

Overall, the present prescription of the sink simulation do not reproduce the evolution of the individual fragment of the stiff-EOS run, but a rough agreement in the evolution of the total mass/number of the fragments is found.
}

\section{Comparison with previous simulations in the literature}
We find a simple power-law growth of the number of fragments from our limited sample of simulations. In addition, there are a number of calculations by various authors, although the assumptions of the simulations are different from each other. At first glance, it seems to be difficult to compare these calculations with each other, but according to the scale-free nature of the system, we can compare the results by scaling the time after the formation of the first protostar. 
The basic equations we solve are
\begin{eqnarray}
\frac{\partial \rho}{\partial t} +\bm{\nabla}\cdot(\rho\bm{v})&=&0 ,   \nonumber\\
\frac{\partial \bm{v}}{\partial t} +(\bm{v}\cdot\bm{\nabla})\bm{v} &=& -\frac{1}{\rho}\bm{\nabla}P - \bm{\nabla}\Phi , \nonumber\\
P=\kappa \rho^{\gamma_{\rm eff}},&\;&\bm{\nabla}^2\Phi = 4\pi G\rho , \nonumber
\end{eqnarray}
where the variables and constants have ordinary meanings.
There are many scales related to the radiative cooling processes as well as to the chemical reactions.
In fact, most of the previous works solve the energy equation with gas cooling and heating functions instead of using the barotropic relation.
However, it is known that the effective polytrope index is $\gamma_{\rm eff}\simeq 1.09$ for primordial gas clouds \citep{omukai98}.

\begin{table*}
\caption{Previous simulations}
\begin{center}
\begin{tabular}{cccc}
  \hline
Reference & 
Method    &
$n_{\rm th}$ &
Remark
\\
\hline
\citet{stacy10} & sink &$10^{12}{\rm cm^{-3}}$ & Cosmological, 1 halo\\  
\citet{clark11a} & sink &$10^{17}{\rm cm^{-3}}$ & Cosmologiclal, 1 halo, 2 snapshots\\
\citet{greif11} & sink &$10^{17}{\rm cm^{-3}}$ & Cosmological, 5 halos\\
\citet{smith11} & sink &$10^{15}{\rm cm^{-3}}$ & Cosmological, 5 halos, two snapshots\\
\citet{greif12} & no approx. &$10^{19}{\rm cm^{-3}}$ & Cosmological, 4 halos, averaged\\
\citet{stacy12} & sink &$10^{12}{\rm cm^{-3}}$ & Cosmological, 1 halo, RF/NF \\
\citet{susa13} & sink &$3\times 10^{13}{\rm cm^{-3}}$ & BE-sphere, 1 cloud, RF/NF\\
\citet{vorobyov13} & 1 sink + stiff EOS &$10^{14}{\rm cm^{-3}}$ & Cosmological, 1 halo, 2D, time averaged\\
\citet{susa14} & sink &$3\times 10^{13}{\rm cm^{-3}}$ & Cosmological, 59 halos, RF, averaged\\
\citet{machida15} & stiff EOS &$10^{19}{\rm cm^{-3}}$ & BE-sphere, 1 cloud, time averaged\\
\citet{hartwig15b} & sink &$10^{17}{\rm cm^{-3}}$ & Cosmological, 4 halos\\ 
\citet{hosokawa16} & cut cooling &$10^{10}-10^{12}{\rm cm^{-3}}$ & Cosmological, 5 halos, RF/NF, polar coord.\\
\citet{stacy16} & sink &$10^{15}{\rm cm^{-3}}$ & Cosmological, 1 halo, RF\\
\citet{hirano17} & cut cooling &$10^{10},10^{13},10^{15}{\rm cm^{-3}}$ & Cosmological, 1 halo \\
\hline
\end{tabular}
\end{center}
\tablecomments{"RF" and "NF" means the simulations with/without UV radiative feedback. "averaged" denotes that the number in Fig.\ref{fig:num_past} is the averaged number over halos. "time averaged" means the provided time evolution in the reference is time averaged to put on the figure.}
\end{table*}
\label{tab:previous}
\begin{figure*}
	\centering
	{\includegraphics[width=15cm]{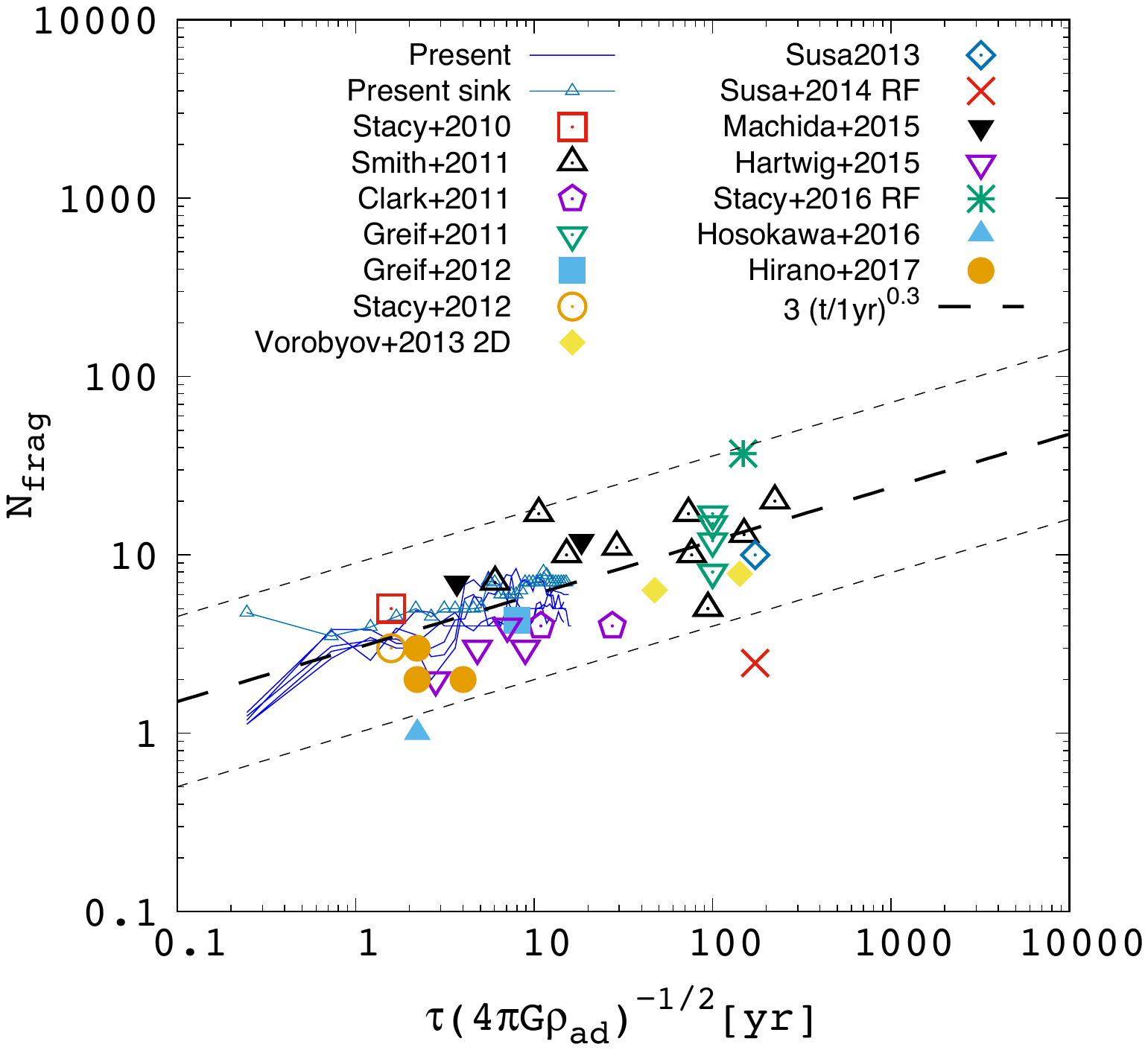}}
	\caption{Number of fragments/sinks vs. scaled time. The scaled time is defined as $\tau/\sqrt{4\pi G\rho_{\rm ad}}$. The thick dashed line shows the fit $\propto t^{0.3}$, and the thin dashed lines are guides for the eyes that denote $\times 3$ and $\times 1/3$ of the thick dashed line. Thin blue lines denote the present results smoothed over {50} yrs as in Fig.\ref{fig:nfrag}, and small triangles with a thin line denote the results of the sink simulation in this work. Open symbols denote the sink simulations \citep{stacy10,clark11a,smith11,greif11,susa13,hartwig15b} while the filled marks are for non-sink simulations \citep{greif12,vorobyov13,hosokawa16,hirano17}. For the case of \cite{vorobyov13} we assume $n_{\rm th}=10^{14}~{\rm cm}^{-3}$, which corresponds to the Jeans length = 6 AU, and the data points are averaged over the first and second 30 kyrs.
    For \cite{hosokawa16}, $n_{\rm th}=10^{10}~{\rm cm}^{-3}$ is assumed.
	The others show results with radiative feedback \citep{stacy12,stacy16,susa14}. For \citet{susa14}, the number is taken from the final time, and the number of fragments are averaged over all the runs.
For the studies that include both of the case with/without radiative feedback, the results from the calculations without radiative feedback are chosen.
{We also tried to plot the results from \cite{riaz18}, but the elapsed time since the first sink formation is unclear, thereby it is omitted. }
}
	\label{fig:num_past}
\end{figure*}

These equations have six variables: $t,\bm{r}, \rho, \bm{v}, P$, and $\Phi$. These quantities can be replaced by the normalization constant and the non-dimensional variables, as: 
\begin{eqnarray}
t&=&t_0\tau,~ \bm{r}=r_0\bm{\xi},~\rho=\rho_0\eta,\nonumber\\
\bm{v}&=&v_0\bm{\zeta},~ P=P_0\sigma,~ \Phi=\Phi_0\phi. \nonumber
\end{eqnarray}
Then, the basic equations are rewritten only with the non-
dimensional variables, as:
\begin{eqnarray}
\frac{\partial \eta}{\partial \tau} +\bm{\nabla}_\xi\cdot(\eta\bm{\zeta})&=&0 , \nonumber\\
\frac{\partial \bm{\zeta}}{\partial \tau} +(\bm{\zeta}\cdot\bm{\nabla}_\xi)\bm{\zeta} &=& -\frac{1}{\eta}\bm{\nabla}_\xi \sigma - \bm{\nabla}_\xi\phi , \nonumber\\
\sigma=\eta^{\gamma_{\rm eff}}~,&\;&\bm{\nabla}_\xi^2\phi = \eta ,  \nonumber
\end{eqnarray}
if the normalization constants satisfy the following five set of equations:
\begin{eqnarray}
t_0&=&\frac{1}{\sqrt{4\pi G\rho_0}},\; v_0=\sqrt{\kappa \rho_0^
{\gamma_{\rm eff}-1}} , \nonumber \\
r_0&=&v_0t_0,\;\Phi_0=v_0^2,\;P_0=\kappa\rho_0^{\gamma_{\rm eff}} . \label{eq:scale} 
\end{eqnarray}
Thus, we have six constants for normalization and five equations. As a result, one constant remains free.
Suppose that we have two simulations with different threshold densities $\rho_{\rm th} (=\mu m_{\rm H}n_{\rm th})$ that characterize the sink formation density or the stiffening of the equation of state. {We can choose this threshold density as the one free normalization constant in place of $\rho_0$.} In that case, if the initial condition for the non-dimensional equation is same, the numerical results should be equivalent {regarding the scaled variables}. In fact, at the onset of the mass accretion phase, the system has converged to the similarity solution, which is almost the same in both of the calculations in terms of the non-dimensional quantities, although some differences could arise from the difference of the numerical methods/implementations or initial conditions. Hence, two distinct calculations with different $\rho_{\rm th}$ should be similar to each other if physical variables are scaled according to the equations (\ref{eq:scale}) with $\rho_0$ replaced by $\rho_{\rm th}$.
One concern is that sink particle simulations normally introduce not only the threshold density but also 
the accretion radius. 
However, 
the accretion radii $r_{\rm acc}$ are chosen so that they are comparable to, or slightly larger than, the Jeans length estimated at the threshold density $\rho_{\rm th}$. Considering that the length is scaled by the Jeans length, as in equations (\ref{eq:scale}), the effect due to the choice of $r_{\rm acc}$ seems limited.

{Table \ref{tab:previous} is the list of numerical simulations in this context so far. In the table, numerical method, threshold density and other specific information are shown.} Fig.\ref{fig:num_past} shows the number of fragments as a function of time scaled by the free-fall time of the threshold density. Simulations in the literature {(\ref{tab:previous})} are shown on the plot with different symbols. We choose the threshold number density {$n_{\rm ad}=10^{19}~{\rm cm^{-3}}$} as the standard, which is the critical density above which the dense core becomes physically adiabatic in the highest resolution simulation \citep{greif12}. Hence, we define the scaled time as $\tau/\sqrt{4\pi G\rho_{\rm ad}}${, where $\rho_{\rm ad} =\mu m_{\rm H}n_{\rm ad}$.} 
Thus, the scaled time of the data point from Fig. 10 in \cite{greif12} is their physical time (8 yrs). 
In other words, the scale time can be regarded as the "real" physical time.

We find that the results from a number of simulations are fundamentally consistent with each other, although they use different numerical methods, initial conditions, cooling functions, and equations of state. In fact, the data points from the cosmological simulation by \cite{smith11} imply that the dependence on the initial conditions already produces the diversity of a factor of $\sim$ 4. 
Thus, this remarkable agreement in this plot tells that the differences caused by the variety of the schemes are at least comparable to the scatter of the results due to the different initial conditions.

We note that the data from \cite{susa14} should be regarded as a lower limit because they took into account the radiative feedback by the protostars, which shut off the fragmentation process in the disk. 
\cite{stacy12,stacy16} also took into account the radiative feedback, but their integrated time is 5000 yrs, which corresponds to the onset of the feedback. Hence, the effect of suppressing the fragmentation is limited. Considering the diversity due to the initial conditions, the results could be compared with the other calculations without radiative feedback.
The data from \cite{hosokawa16} also should be considered as a lower limit, because it used polar coordinates, which tend to have less resolution at the outer part of the disk. The lower resolution results in a smaller fragmentation process in the disk.
\section{Discussion}
We can extrapolate the relation $N_{\rm frag}\propto t^{0.3}$ to several thousand years, the beginning of the radiative feedback from massive protostars. We find the number of fragments is $10$ $\sim$$100$ at that epoch. This number is somewhat larger than expected, especially from  non-sink simulations. One possible reason for this is that non-sink simulations normally follow the evolution of the system for a shorter time than the sink simulations, so they predict fewer fragments. If we regard the feedback as strong enough to shut off the disk fragmentation/merger, we arrive at the final number of the protostars at that epoch {i.e. 10-100. 
 
 However, the previous studies \citep{hosokawa11,hosokawa16,susa13,stacy16} predict that the effects of radiative feedback becomes prominent $10^3-10^4$ yrs after the formation of the first protostar, not in good agreement. If the photoevaporation of the disk proceeds slowly and the fragmentation/merger processes continue to much later time, the expected number of fragments could be different. It is a disk fragmentation problem under UV radiation field, in which the outcome do not scales by the relation discussed in the previous section. Thus, it has to be investigated in the future studies.}
 

{In any case,} we have to extend our non-sink/sink simulations {\it without} UV feedback to several thousand years in scaled time to confirm the $N_{\rm frag}\propto t ^{0.3}$ trend all the way to the onset of the radiative feedback. Then, the simulations should be followed by calculations with UV feedback to obtain the final mass distributions, binary frequencies, and so forth. 

{Another complexity is coming from the magnetic field associated with the turbulence. 
Very high resolution simulations\citep{sur12,federrath11,turk12} as well as the analytic calculations using Kazantsev equation\citep{schleicher10,schober12} show that the minihalo is turbulent and the initial seed magnetic field is easily amplified to the equipartition level. Because of the strong coupling between the magnetic field and the gas in primordial gas\citep{maki04,maki07,susa15,higuchi18}, B-field is not dissipated at the Jeans scale, to be amplified efficiently. The effects of magnetic field on the PopIII star formation has been investigated by \cite{machida_doi13}, where they find that the  circumstellar disk formation is suppressed in the presence of equipartition level B-field, because of the efficient angular momentum transportation by the magnetic breaking. As a result, no fragments are found. However they assume a coherent field parallel to the rotation axis, which is quite different from the turbulent magnetic field expected in the minihalos\footnote{\citet{latif16} discussed the $\alpha\Omega$ dynamo process in the accretion disk, where they find the coherent field could be generated from the turbulent field.}. On the other hand \cite{seifried12} reported that turbulence can circumvent the magnetic breaking catastrophe in the context of the present-day star formation. Thus, the effects of turbulent magnetic field on the PopIII disk fragmentation is still an open question, which should be addressed in the future.}

\section{Summary}
We perform cloud collapse simulations with a barotropic equation of state derived from the one-zone model of the gravitationally contracting primordial gas cloud, to mimic the formation of PopIII stars. 
We find growing disk-like structures after the formation of the first fragment, followed by the rapid fragmentation of the disk and the merger of the fragments when the disk becomes massive enough to be gravitationally unstable.
We find that most of the fragmentation events occur in the accretion disk around individual fragments.
These results suggest a simple analytical model of the evolution of the number of fragments that predicts a power-law growth of them. In fact, we find that the number of fragments slowly increases with time, following the relation $N_{\rm frag}\propto t^{0.3}$. 
{We also perform a simulation with standard sink particles, where the number and total mass of sink particles are in rough agreement with those of the stiff equation of state runs.}
Finally, we compare the number of fragments with other published results, by scaling the simulated time according to the notion of the scale-free nature of the system. Consequently, we find a good agreement among most of the calculations so far. 
The present results {combined with the studies in the literature imply} that the population III stars are not born as single stars, but in a multiple system.

\hspace{1cm}

\bigskip
HS thanks {the anonymous referee for careful reading and constructive comments.}, T. Hosokawa, M. Machida, G. Chiaki, S. Hirano and T. Hartwig for fruitful discussions. We thank the support by Ministry of Education, Science, Sports and Culture, Grant-in-Aid for Scientific Research Nos. 17H02869,  17H01101, and 17H06360. 


\end{document}